\documentclass[ 
                reprint, 
                superscriptaddress, 
                aps, 
                prl, 
                citeautoscript]{revtex4-2}
\usepackage{graphicx}
\usepackage{dcolumn}
\usepackage{bm}
\usepackage{amsmath,amssymb}
\usepackage[version=4]{mhchem}
\usepackage{xcolor}
\usepackage{siunitx}
\DeclareSIUnit\angstrom{\text {Å}}
\usepackage{braket}
\usepackage{xr}
\usepackage{float}
\usepackage{multirow}
\usepackage{wrapfig}
\usepackage{gensymb}
\usepackage{quotes}
\usepackage{xr}
\nocite{*}

\begin{document}






\title{Nonlinear Dynamics in the Formation of Molecular Polariton Condensates}


\author{Evan~J~Kumar}
\affiliation{Department of Physics \& Center for Functional Materials, Wake Forest University, Winston-Salem, North Carolina 27109, United~States}

\author{Katherine~A~Koch}
\affiliation{Department of Physics \& Center for Functional Materials, Wake Forest University, Winston-Salem, North Carolina 27109, United~States}

\author{Rishabh Kaurav}
\affiliation{Department of Physics, Center for Discovery and Innovation, The City College of New York, 85 St. Nicholas Terrace, New York, New York 10031, United States}

\author{Ravindra Kumar Yadav}

\affiliation{School of Physical Sciences, Indian Institute of Technology Mandi, Mandi 175005, Himachal Pradesh, India}

\author{Victoria~Quir\'os-Cordero}
\affiliation{School of Materials Science and Engineering, Georgia Institute of Technology, 771 Ferst Dr NW, Atlanta, GA~30332, United~States}

\author{Josiah~N~Brinson}
\affiliation{Department of Physics \& Center for Functional Materials, Wake Forest University, Winston-Salem, North Carolina 27109, United~States}

\author{Vinod~Menon}
\email{vmenon@ccny.cuny.edu}
\affiliation{Department of Physics, Center for Discovery and Innovation, The City College of New York, 85 St. Nicholas Terrace, New York, New York 10031, United States}

\author{Ajay~Ram~Srimath~Kandada}
\email{srimatar@wfu.edu}
\affiliation{Department of Physics \& Center for Functional Materials, Wake Forest University, Winston-Salem, North Carolina 27109, United~States}

\date{\today}

\begin{abstract}
    Exciton-polaritons—hybrid light-matter quasiparticles—can undergo Bose-Einstein-like condensation at elevated temperatures owing to their lower effective mass. This becomes even more pronounced in the context of molecular polariton condensates where the large exciton binding energy of Frenkel excitons facilitates condensation at room temperature. While widely studied as low-threshold coherent light sources, the dynamics of their condensation remain poorly understood, partly due to the limitations of existing kinetic models. Here, we use excitation correlation photoluminescence (ECPL), a nonlinear optical technique with 220\,fs resolution, to probe molecular polariton condensation in Rhodamine-B-doped small-molecule ionic isolation lattices (SMILES). This platform promotes dipole alignment and suppresses detrimental intermolecular interactions. ECPL reveals condensate formation within hundreds of femtoseconds, driven by radiative scattering from the reservoir. A sustained population beyond the polariton lifetime suggests an additional feeding mechanism from higher momentum states in the lower polariton dispersion. These results provide quantitative insight into condensation timescales and mechanisms, advancing our control over polariton dynamics.

\end{abstract}
 
\maketitle


Exciton-polaritons are hybrid light-matter quasiparticles formed via strong coupling between material excitations and confined cavity photons. Due to their low effective mass, polariton undergo condensation similar to Bose-Einstein condensation (BEC) at relatively low densities and elevated temperatures, making them promising candidates for low-threshold coherent light sources~\cite{deng2003polariton,deng2010exciton}. These out-of-equilibrium condensates have been demonstrated in a range of systems, including inorganic quantum wells and organic chromophores~\cite{keeling2020bose,daskalakis2014nonlinear,kasprzak2006bose,kena2010room}. While large exciton binding energy, strong oscillator strength, and narrow emission linewidths are necessary material properties~\cite{wang2018colloquium,bittner2012estimating}, they are not sufficient to ensure efficient polariton accumulation. A comprehensive understanding of how material and photonic factors determine condensation dynamics and threshold densities remains lacking, in part due to variations in the photophysical mechanisms that govern polariton relaxation across different systems~\cite{kavokin2022polariton,sanvitto2016road}.

In organic polariton systems, strong coupling generates a dense manifold of dark excitonic states (the reservoir) in addition to the upper and lower polariton branches~\cite{dong2022dynamics}. Condensation is driven by population transfer from the reservoir and high-$k$ polaritons into the lower polariton ground state at $\vec{k}=0$, a process competing with finite polariton and reservoir lifetimes~\cite{mazza2013microscopic}. These dynamics are typically described using semiclassical rate equations or open-dissipative Gross-Pitaevskii models~\cite{wouters2007excitations,keeling2008spontaneous}, both of which rely on multiple phenomenological parameters. However, parameter extraction is often ambiguous, particularly when based on steady-state emission or time-resolved photoluminescence alone.

Time-resolved absorption and reflection techniques can offer more direct insight into polariton population dynamics~\cite{yamashita2018ultrafast,ramezani2019ultrafast}, but their application in high-$Q$ cavities is hindered by low signal transmission and the mixed optical response of the cavity~\cite{ashoka2022extracting}. Moreover, time-resolved PL is complicated by overlapping signals from the reservoir and polariton states, limiting its ability to directly resolve condensate formation. As a result, few techniques can access sub-picosecond condensation dynamics in molecular polariton systems with both specificity and temporal resolution.

Here, we employ excitation correlation photoluminescence (ECPL), a nonlinear optical technique with $\sim$220\,fs temporal resolution~\cite{srimath2016nonlinear,rojas2023resolving}, to probe the dynamics of polariton condensation in Rhodamine-B (R3B) embedded in small-molecule ionic isolation lattices (SMILES)~\cite{deshmukh2024plug}. The SMILES framework suppresses aggregation and nonradiative loss pathways, providing a clean model system. ECPL measurements reveal sub-picosecond condensate formation following off-resonant excitation, driven by radiative scattering from the excitonic reservoir. Notably, the observed growth occurs faster than the cavity photon lifetime, and ECPL dynamics indicate an additional population channel that sustains condensate occupation beyond polariton lifetimes. Complementary steady-state and transient measurements support the role of non-negligible contributions from the population at higher momenta in this process. Together, these findings provide direct insight into the timescales and mechanisms underlying molecular polariton condensation.

\begin{figure}[t!]
    \centering
    \includegraphics[width=3.375in]{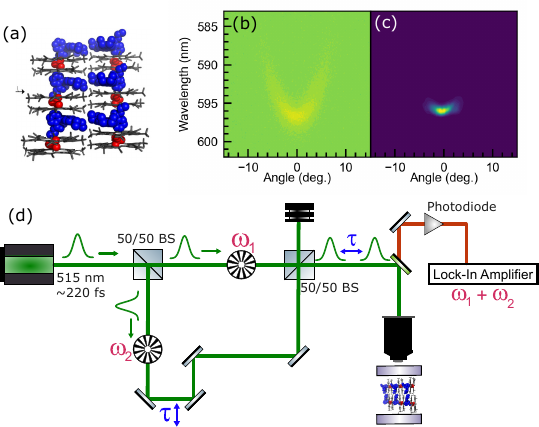}
    \caption{(a) Chemical structure of R3B-SMILES. (b) and (c) angle resolved PL maps at below and above condensation thresholds. Scheme of the experimental setup used to measure the ECPL dynamics: BS is a beam splitter, $\omega_1$ and $\omega_2$ are the two modulation frequencies and $\tau$ is the time delay betweeen the pulses.} 
    \label{fig:setup}
    \vspace{-20pt}
\end{figure}

To investigate the ultrafast dynamics of polariton condensate formation and decay, we employ a strongly coupled microcavity incorporating a rhodamine-based chromophore system as the excitonic medium (see SI Section I for sample preparation \cite{SIfile}). Specifically, we use R3B-SMILES—whose molecular structure is shown in Fig.~\ref{fig:setup}(a)—a system demonstrated to support polariton condensation. This behavior is attributed to the chemical scaffold of R3B-SMILES, which spatially isolates individual chromophores and mitigates detrimental intermolecular interactions \cite{benson2020plug}. Figs.~\ref{fig:setup}(b) and (c) present angle-resolved photoluminescence spectra under non-resonant excitation at 515\,nm, recorded below and above the condensation threshold, respectively. The data in Fig.~\ref{fig:setup} confirm the formation of a polariton condensate at room temperature.

We measure the ultrafast dynamics of the condensate using the ECPL technique. The experiment involves photo-exciting the sample of interest with two identical pump pulses that are temporally displaced from each other and measuring the nonlinear photoluminescence (PL) at distinct temporal delays between them as shown in Fig.~\ref{fig:setup} (see SI section I for details \cite{SIfile}. 
The emission from the microcavity is collected under normal incidence, thus close to the $\vec{k} = 0$ point in the dispersion, and detected using a slow photodiode. The time-integrated PL at the photodiode can be estimated as: 

\begin{multline}
\label{ecpl}
PL(2I_{pump},\tau) = \int_{0}^{\infty} PL_{1}(I_{pump},t)dt\\
+ \int_{\tau}^{\infty} PL_{2}(I_{pump},t)dt + \int_{\tau}^{\infty} PL_{cross}(2I_{pump},t,t-\tau)dt,
\end{multline}

Here, $PL_1$ and $PL_2$ denote the photoluminescence (PL) contributions from pump pulses 1 and 2, respectively, and $\tau$ is their temporal delay. The total PL includes contributions from each pulse and a cross-correlation term (Eq.~1), indicative of nonlinear population mixing. The ECPL response, defined as $\Delta PL/PL$, quantifies this nonlinear contribution, isolated experimentally by demodulating the detector signal at the sum frequency $\omega_1+\omega_2$ using a lock-in amplifier (see section I of SI \cite{SIfile}). ECPL thus serves as a sensitive probe of PL intensity variations with excitation density. For linear PL scaling, the ECPL signal vanishes; deviations yield a nonzero response---positive for super-linear, negative for sub-linear scaling~\cite{rojas2023resolving}.

At high excitation densities, non-radiative exciton annihilation in organic films~\cite{dexter1954theory,lunt2009exciton,akselrod2010exciton} leads to sub-linear PL and negative ECPL. For R3B-SMILES films, we observe such negative ECPL signals persisting over hundreds of picoseconds (Fig.S3 \cite{SIfile}), increasing in magnitude with excitation density. This behavior reflects nonlinear quenching dynamics, from which we estimate an exciton lifetime of $\sim$3\,ns.

\begin{figure}[t!]
    \centering
    \includegraphics[width=3.3in]{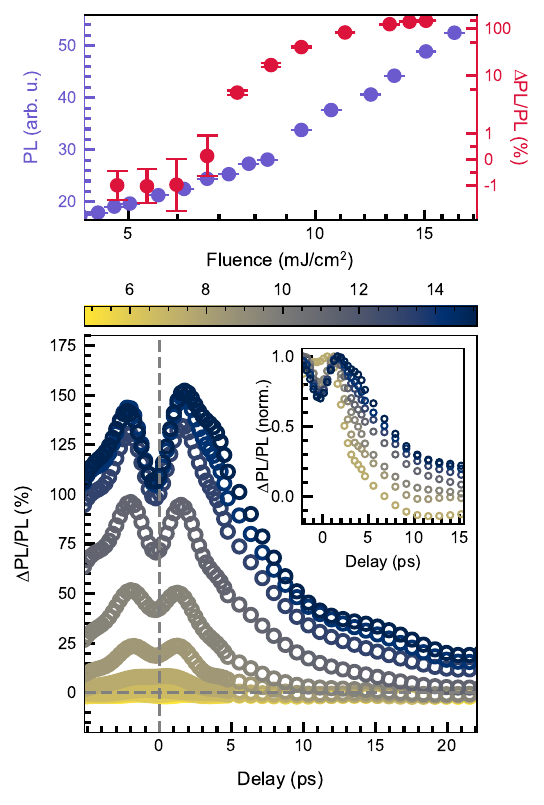}
    \caption{(a) Experimental plot of the total integrated photoluminescence (purple) and the $\%\Delta$PL/PL (red) as a function of excitation fluence. The signals show a super-linear behavior above a threshold fluence of 8 mJ/cm$^{2}$. (b) Experimental time-resolved ECPL dynamics showing two distinct timescales as the fluence is increased across the threshold. The inset shows the normalized ECPL dynamics for comparison of the timescales across the different fluences.}
    \label{fig:exp}
\end{figure}

We now examine the photoexcitation dynamics in R3B-SMILES microcavities using the ECPL technique. Figure~\ref{fig:exp}(a) shows a log-linear plot of lower PL intensity under single-pulse excitation at 515\,nm, integrated spectrally and plotted against pump fluence (purple). A distinct slope change near $\sim$8 mJ/cm$^2$ marks the threshold for polariton condensation, characterized by a nonlinear PL increase due to bosonic stimulation \cite{kasprzak2006bose}. This transition is further supported by k-space PL data in Fig.~\ref{fig:setup}(c). Sample degradation is observed at threshold, with unknown origin; to mitigate this, the excitation spot is continuously moved during measurements (see section III of SI \cite{SIfile}).


Figure~\ref{fig:exp}(a) also presents a log-log plot of the relative ECPL signal, $\Delta$PL/PL, measured at a 2.5\,ps inter-pulse delay (red). The fluence refers to the combined energy of both excitation pulses. Below threshold, the signal is near zero, indicating minimal nonlinear response. At $\sim$8\,mJ/cm$^2$, a sharp increase in ECPL is observed, becoming strongly positive and peaking near $\sim$11\,mJ/cm$^2$ before gradually tapering off. This positive ECPL response reflects the super-linear PL growth driven by bosonic final-state stimulation at the $\Vec{k} = 0$ lower polariton state \cite{waldherr2018observation,zasedatelev2021single}.


The PL from the microcavity exhibits an intrinsic negative nonlinearity due to non-radiative quenching of the exciton reservoir population, driven by bimolecular interactions—an effect also observed in the weak coupling regime in thin films. While this negative contribution persists across all excitation fluences, beyond the condensation threshold, it is overshadowed by the strong positive nonlinearity arising from the emission of the polariton condensate. This transition is reflected in the data presented in Fig.~\ref{fig:exp}(b). Consequently, we identify the emergence of a positive ECPL signal as a defining characteristic of the polariton condensate. The ECPL dynamics, depicted in Fig.~\ref{fig:exp}(b), thus serve as a direct representation of the condensate formation and evolution, providing valuable insight into its underlying kinetics.

We observe that the ECPL dynamics remain negative below the condensation threshold, evolving on the hundreds of picoseconds timescale. This behavior is governed by the exciton reservoir dynamics and closely resembles the ECPL response observed in thin films and low-Q cavities that do not support condensation (see Fig. S7 \cite{SIfile}). However, when the total excitation fluence surpasses 7.49\,mJ/cm$^2$, a positive ECPL signal emerges within 300–500\,fs, followed by a decay over a few picoseconds.

Given that a positive ECPL signal is characteristic of polariton condensation, we infer that the condensate forms on a sub-picosecond timescale and has a lifetime of only a few picoseconds. Notably, across all fluences shown in Fig.~\ref{fig:exp}(b)--except for the highest fluence of 15.08\,mJ/cm$^2$—the individual pulse fluences remain too low to independently generate a condensate. Nevertheless, the initial pulse creates a reservoir population that persists long enough to interact with the second pulse’s population, collectively sustaining the positive ECPL signal and, consequently, the condensate. This observation underscores the role of long-lived reservoir interactions in facilitating condensate formation under pulsed excitation.

Polariton condensate dynamics have been modeled using various theoretical approaches. Mean-field treatments based on the Gross-Pitaevskii equation effectively capture threshold behavior and fluence dependence, particularly in inorganic systems \cite{wouters2007excitations,mukherjee2019observation}. In organics, however, additional complexity from non-radiative relaxation pathways complicates such modeling. Nonetheless, kinetic rate-equation models tracking incoherent population dynamics can describe condensation. For instance, Mazza et al.~\cite{mazza2013microscopic} employed coupled rate equations for the exciton reservoir, the lower polariton dispersion, and the $\vec{k} = 0$ mode, yielding good agreement with experimental condensation thresholds in anthracene cavities.

\begin{figure}[t!]
    \centering
    \includegraphics[width=3.375in]{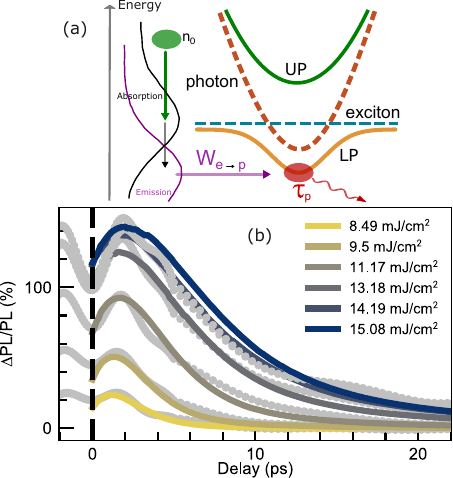}
    \caption{(b) Energetic structure of the R3B-SMILES system. The non-resonant excitation creates a population, $n_{0}$, that thermalizes quickly to the dark state reservoir. The reservoir radiatively pumps and populates the LP branch ($W^{e\rightarrow{}p}$). The emission from the LP happens on a timescale that is related to the polariton lifetime, $\tau_{p}$.  (b) The experimental ECPL dynamics are reproduced in gray for a range of fluences above the condensation threshold. The fitted simulations from the semi-classical Boltzmann equations are plotted over the data (blue-yellow). }
    \label{fig:sim}
\end{figure}

We follow a similar approach to model the measured ECPL dynamics. For simplicity, we only consider the population at $\vec{k}=0$ polariton state along with the exciton density in the reservoir and ignore the population at larger $\vec{k}$ in the LP dispersion. This is consistent with the previous consideration of radiative pumping directly into the $\vec{k}=0$ state being the dominant mechanism driving the condensation process. In our picture, the non-resonantly generated population in the reservoir relaxes, via internal conversion, within the time resolution of the experiment into the lowest excited state of the molecule. Thus, we ignore the pumping term and assume that the population is directly injected into the lowest state in the reservoir. Part of the photo-generated population is transferred into the polariton state through the radiative pumping mechanism, see the scheme in Fig.~\ref{fig:sim}(b). Accordingly the evolution of the exciton population ($n_e$) and the population in the $\vec{k}=0$ polariton state ($n_p$) can be written as:
\vspace{-10pt}
\begin{align}
\dot{n}_e &= -\Gamma_e n_e - W^{e\rightarrow p} n_e\left(1 + \frac{n_p}{\Tilde{n}_p} \right) \nonumber\\
&\quad - \gamma'(n_e + |c_p^{(e)}|^2 n_p) n_e, \nonumber\\
\dot{n}_p &= -\Gamma_p n_p + W^{e\rightarrow p} n_e\left(1 + \frac{n_p}{\Tilde{n}_p} \right) \nonumber\\
&\quad - \gamma'(n_e + |c_p^{(e)}|^2 n_p) |c_p^{(e)}|^2 n_p.
\label{eq:rates}
\end{align}



Here, $\Gamma_e =1/\tau_e$ and $\Gamma_p = 1/\tau_p$ are the exciton and polariton lifetimes, respectively, and $\gamma'$ represents the bimolecular quenching term. The latter accounts for the bimolecular quenching of the polariton population due to the interactions with the reservoir as well as the background polaritons. These interactions are considered to be mediated by the excitonic component of the polaritons \cite{estrecho2019direct}, and accordingly the associated rates are scaled by $|c_p^{(e)}|^2$, the excitonic Hopfield coefficient (see Fig. S2 in SI \cite{SIfile}). The bimolecular quenching can also non-radiatively annihilate the population in the reservoir. The critical component of the model is the term with the radiative pumping rate, effectively captured in a single phenomenological parameter $W^{e\rightarrow{}p}n_{e}$. This term represents stimulated scattering of the exciton population into the condensate when the polariton population exceeds the threshold density defined by ${\Tilde{n}_{p}}$. The initial excitation density of the pump, $n_{0}$, can be calculated from information from the cavity, the excitation source, and ECPL measurements of the bare R3B-SMILES film (shown in Fig. S2 in SI).

\raggedbottom

We numerically solve the rate equations to determine each population as a function of time as we impose the initial condition: $t=0$, $n_{e} = n_{0}$. Integrating the population of the LP branch at $\vec{k}=0$ over all times, we compute a number that is proportional to the total photoluminescence signal. To simulate the ECPL experiment, we inject another identical population into the reservoir at each delay and integrate to get the total photoluminescence signal. By subtracting twice the value of the PL from single excitation pulse, we determine a value for $\Delta$PL/PL. 

\raggedbottom


In Fig.~\ref{fig:sim}(a), we depict the energetic landscape of the system and show how the system relates to various parameters in our rate equations. In Fig.~\ref{fig:sim}(b), we reproduce the plots of the experimental data (gray dotted) and plot the best-fit curves from the semiclassical Boltzmann equations on top (blue-yellow), whose parameters are determined by a least-squares regression algorithm. The fitted curves successfully capture the growth and decay dynamics observed in the experiment, though some deviations become more pronounced at higher excitation densities. We observe that the estimated rate coefficients—$W^{e\rightarrow{}p}$, $\gamma'$, and $\tau_p$—needed to be adjusted within a range across different excitation fluences to achieve reliable fits. Specifically, $W^{e\rightarrow{}p}$ is found to lie between 1000–5000,s$^{-1}$, $\tau_p$ between 0.5–1.2,ps, and $\gamma'$ between 0.1–1$\times$10$^{10}$,cm$^2$s$^{-1}$ (see SI Fig. S1 \cite{SIfile}). These values are consistent with those previously reported for molecular polariton condensates~\cite{daskalakis2014nonlinear, mazza2013microscopic}.

Although the variation in fitting parameters remains within an order of magnitude—likely due to numerical uncertainties, experimental variability, or inaccuracies in estimating the initial excitation density—we emphasize that features in the experimental data may also drive these changes. In particular, the ECPL dynamics exhibit slower formation and decay rates at higher excitation fluences. This behavior is clearly visible in the normalized dynamics shown in Fig.~\ref{fig:exp}(b), where the ECPL signal persists longer with increasing fluence.

The fitting algorithm reflects this trend by systematically reducing the rate coefficients to match the experimentally observed slowdown (see Fig. S1 \cite{SIfile}). We interpret this variation not merely as a fitting artifact, but as evidence of genuine changes in the system’s dynamics. Notably, the variation in parameters lacks a clear physical justification within a given microcavity design, which strongly points to limitations in the kinetic model's ability to describe the condensate behavior—particularly under high excitation fluence. For fluences exceeding twice that threshold, the extended dynamics may reflect many-body effects or interaction-induced bottlenecks. However, slow dynamics are also observed at modest fluences, around 1.1–1.3 times the threshold, suggesting the presence of an intrinsic dynamical process not captured by the current model.   

\begin{figure}[t!]
    \centering
    \includegraphics[width=3.37in]{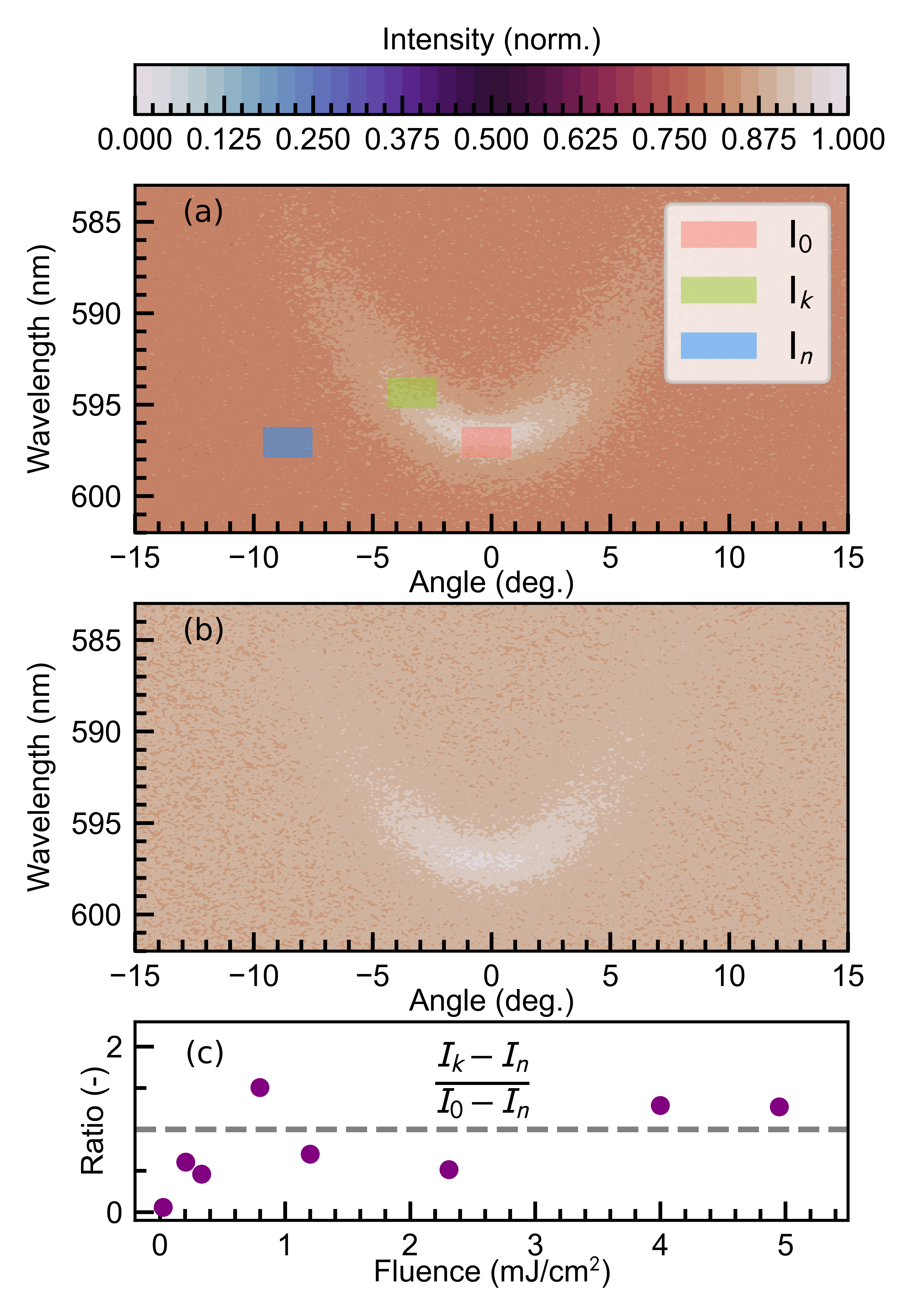}
    \caption{$k$-space resolved photoluminescence map of R3B-SMILES cavity at excitation fluence of (a) 2.5 and (b) 0.7\,mJ/cm$^2$. (c) Fluence dependence of the ratio of photoluminescence intensity at 3$^\circ$-5$^\circ$ and PL intensity at 0$^\circ$ with the background subtracted, indicating an increased realtive density at higher-lying $k$-states.}
    \label{fig:Kspace}
    \vspace{-15pt}
\end{figure}

A likely explanation is the occupation of higher $\vec{k}$-states above $\vec{k}=0$, not included in the current model. Given the broad R3B-SMILES emission, population of higher $\vec{k}$-states is expected even though the cavity is designed to favor $\vec{k}=0$. However, simple considerations do not predict a fluence-dependent increase in such a population. Direct evidence for higher $\vec{k}$ occupation is found in $k$-space-resolved PL maps below threshold [Figs.~\ref{fig:Kspace}(a),(b)]. At both 2.5 and 0.7 mJ/cm$^2$ fluences, emission across the lower polariton branch is observed, consistent with broad radiative pumping. Notably, the relative intensity of higher $\vec{k}$-states increases with fluence, with each map normalized to the $\vec{k}=0$ peak.



To quantify this, we define regions of interest (ROIs) for $\vec{k}=0$ ($I_0$), 3°–5° ($I_k$), and background ($I_n$) [Fig.\ref{fig:Kspace}(a)], and compute $(I_k - I_n)/(I_0 - I_n)$. The resulting ratios, plotted in Fig.\ref{fig:Kspace}(c), increase systematically with fluence, confirming the growing occupation of higher $\vec{k}$-states.


Transient reflectivity measurements (see SI section V \cite{SIfile}) provide further evidence of population buildup in higher $k$-states. Using low-quality-factor cavities to suppress top-mirror contributions, we observe a pronounced bleach feature emerging at 600--610\,nm within a few picoseconds, consistent with the slow population of the lower polariton state. These timescales match those from ECPL measurements, indicating that condensate formation is largely governed by the radiative pumping rate. In higher-$Q$ cavities, as used in ECPL, the peak intensity is expected to be reached more rapidly.

At later times ($\sim$4\,ps), once the growth dynamics are complete, the bleach peak exhibits a fluence-dependent blue shift, while the dynamics remain largely unaffected. This shift reflects increased occupation of higher-energy $k$-states. Although the setup lacks $k$-space resolution, the measurement geometry captures population changes across a broad range near $\vec{k} = 0$, with higher-$k$ population manifesting as a blue shift in the $k$-integrated signal. While refractive index changes at high excitation could also induce blue shifts, such effects typically produce derivative-like features, which are not observed here.

By combining the fluence dependence observed in $k$-space-resolved PL spectra and transient reflectivity measurements, we hypothesize that higher-lying $k$-states increasingly contribute to polariton dynamics at elevated fluences, particularly near the condensation threshold. This redistribution of population across the dispersion may be driven by pair-scattering processes that depopulate the zero-momentum state. Such mechanisms are generally unexpected in polaritons based on Frenkel excitons, due to their weaker excitonic interactions, especially near $\vec{k} = 0$ \cite{yagafarov2020mechanisms}. Alternatively, saturation effects at high excitation densities may suppress population buildup at low momenta. Further experiments, involving systematic variation of cavity coupling parameters and complementary theoretical modeling, are needed to test these interpretations and clarify the role of momentum-dependent dynamics in the condensation process~\cite{balasubrahmaniyam2023enhanced, michail2024addressing}. In particular, estimating the $k$-dependence of the ECPL and transient reflectivity responses will be critical.

In summary, we employ excitation correlation photoluminescence to obtain an unambiguous measurement of the formation and decay dynamics of molecular polariton condensates. Our measurements reveal that the condensate formation occurs on sub-picosecond timescales via a radiative scattering mechanism from the exciton reservoir. A kinetic model incorporating radiative pumping and bosonic stimulation successfully reproduces the ECPL dynamics but does not fully capture the density-dependent trends observed in experiments. Complementary steady-state and transient measurements suggest that higher-energy state scattering may play a role in these discrepancies, underscoring the need for refined theoretical models. These findings provide insight into the mechanisms governing polariton condensation dynamics and highlight the capabilities of ECPL as a tool for probing ultrafast photophysical processes in microcavity systems.

\section{Acknowledgements}
ARSK acknowledges funding from the National Science Foundation CAREER grant (CHE-2338663), start-up funds from Wake Forest University, funding from the Center for Functional Materials at Wake Forest University. Work at City College of New York was supported by the U.S. Air Force Office of Scientific Research MURI
Grant FA9550-22-1-0317 (R.K) and the National Science Foundation grant OMA-2328993 (V.M.M). Any opinions, findings, and conclusions or recommendations expressed in this material are those of the authors(s) and do not necessarily reflect the views of the National Science Foundation.

\section{Data Availability Statement}
The data that support the findings of this article are openly available.~\cite{kumardata25}

\section{References}
%




\end{document}